\documentstyle[aps,preprint]{revtex} 
  \begin{document} 
  %@twocolumnfalse\endcsname 
  %\setlength{\baselineskip}{22pt} 
  \draft 
  %\preprint{} 
  \title{Enhancement of the Kondo temperature of 
  magnetic impurities in metallic point contacts due to the 
  fluctuations of the local density of states} 
  \author{G. Zar\'and and L. Udvardi} 
  \address{Institute of Physics, Technical University of 
  Budapest, \\ 
  H 1521 Budafoki \'ut 8., Budapest, Hungary.} 
  \date{\today} 
  \maketitle 
  \begin{abstract} 
  The effect of local density of states (LDOS) fluctuations on the 
  dynamics of a Kondo impurities in a small metallic point contact 
  (PC) is studied. 
  To estimate the spatial and energy dependent LDOS fluctuations 
  we investigate a model PC 
  by means of  a transfer matrix formalism. 
  For small PC's in the nanometer scale we find that near to 
   the orifice strong LDOS fluctuations develop. 
  These fluctuations  may 
  shift the Kondo temperature by several orders of 
  magnitude, and result in a strong broadening of the PC Kondo 
  peak 
  in agreement with the results of recent measurements. 
  \end{abstract} 
  %\vskip2pc] 

  Magnetic impurities have been studied in the past few years very 
  successfully using ballistic point contacts (PC's) 
  \cite{Lysykh,Omel,Jansen,Naidyuk}. 
  Very recently a thorough reinvestigation of the Kondo effect has 
  been carried out by Yanson {\it et al.} \cite{Yanson} using 
  small metallic 
  PC's with the diameters in the mesoscopic range, $d\sim 20 \AA$. 
  Surprisingly, in contrast to measurements in thin metallic films 
  and quantum wires, where a suppression of the Kondo effect has 
  been found   \cite{Kondo_films}, Yanson {\it 
  et al.} observed a strong broadening of the zero bias Kondo peak 
  which has been explained by the increase of the Kondo 
  temperature of the impurities in the contact region 
  \cite{Yanson}.

  In this Letter we report on the first study of the local 
  density of states (LDOS) and their effect on the Kondo resonance 
  in small ballistic PC's. We find surprisingly 
  strong LDOS fluctuations in the contact region which are 
  generated by scattering of the conduction electrons at the 
  surface of the PC and give a natural explanation 
  to the broadening of the Kondo peak in very small PC's 
  \cite{Yanson}. 
   
  The effect of LDOS fluctuations on the Kondo resonance has 
  already been studied in the case of disordered metals 
  \cite{Dobros}. However, Dobrosavljevi\'c {\it et al.} calculate 
  the effect of the fluctuating LDOS only in the leading 
  logarithmic order and their results are only valid in the case 
  where the characteristic energy scale of the fluctuations, 
  $\epsilon_c$ is of the same order of magnitude as the bandwidth 
  cutoff, $D\sim \epsilon_c$. As we shall see in a PC 
  $\epsilon_c\ll D$ thus we had to treat the fluctuations with 
  more care. Furthermore, we also constructed the next to leading 
  logarithmic scaling equations wich goes beyond the theory of 
  Ref.~\cite{Dobros}. 
   
  To study the effects of the LDOS fluctuations on a Kondo 
  impurity 
  we investigate the dynamics of a magnetic impurity in the 
  contact region. For this purpose we use a simplified 
  Hamiltonian. The first part of our Hamiltonian describes the 
  conduction electrons: 
  $H_0 = \sum_{n,\sigma}\epsilon_n a^+_{n\sigma}a_{n\sigma}$, 
  where the operator $a^+_{n\sigma}$ ($a_{n\sigma}$) creates 
  (annihilates) a conduction electron with spin $\sigma$, wave 
  function $\varphi_n$, and energy $\epsilon_n$. The states 
  $\varphi_n$ are not momentum eigenstates 
  but rather appropriately chosen scattering states of the PC. 
  The interaction part of the Hamiltonian can be written as 
  \begin{equation} 
  H_{\rm int}=\sum_{n, n^\prime, \sigma, \sigma^\prime, i} 
  \varphi_n^\ast({\bf r}) J \varphi_{n^\prime}({\bf r})\; 
  a^+_{n\sigma}\sigma^i_{\sigma\sigma^\prime} a_{n^\prime 
  \sigma^\prime} S^{\; i}\;, 
  \label{eq:H_int} 
  \end{equation} 
  where the impurity is at the position ${\bf r}$, $J$ is the strength 
  of the local exchange interaction, and $\sigma$ and $S$ denote 
  the spin of the conduction electrons and the impurity, 
  respectively.

  Far from the orifice the LDOS must approach its bulk value: 
  $\varrho({\bf r},\epsilon)\to \varrho_{\rm bulk}(\epsilon)$. For the 
  sake of simplicity we assume that the bulk DOS is constant 
  between the high- and low-energy cutoffs, $\varrho_{\rm 
  bulk}(\epsilon) = \varrho_0$. However, approaching the contact 
  region $\varrho({\bf r},\epsilon)$ deviates from its bulk value and 
  random spatial and energy dependent LDOS fluctuations appear. 
   
  To estimate the amplitude of these fluctuations we carried out 
  numerical calculations for a free electron model PC consisting 
  of two infinite half-spaces connected by a cylinder of radius 
  $R$ 
  and length $L$ \cite{Zar_Udv}. To simplify the calculations we 
  used angular 
  momentum eigenstates propagating along the axis $z$ of the PC 
  of the form: 
  \begin{equation} 
  \varphi_{\pm,\epsilon \lambda m}(r,z,\varphi) = e^{i\varphi m} 
  J_m(\lambda r) e^{\pm i k_z(\lambda)\;z} \; , 
  \label{eq:states} 
  \end{equation} 
  where $m$ is the angular momentum around the axis $z$ and 
  cylindrical coordinates have been used. The signs $\pm$ 
  correspond to right- and left-going states, respectively, 
  $\epsilon$ denotes the energy of the conduction electron, and 
  $J_m$ 
  stands for the $m$'th Bessel function. The $z$ component of the 
  momentum in Eq.(\ref{eq:states}), $k_z$, can be expressed by 
  the energy $\epsilon$ and the radial momentum 
  $\lambda$ of the electron as $k_z = (2m_e \epsilon - 
  \lambda^2)^{1/2}$ and $k_z=i(\lambda^2-2m_e\epsilon)^{1/2} $ for 
  $\sqrt{2m_e\epsilon}   
  \lambda$ and $\sqrt{2m_e\epsilon} < \lambda$, respectively 
  ( $m_e$ denotes the electron mass). Since we require the 
  vanishing of the wave functions at the boundaries, $\lambda$ is 
  a continuous parameter in the infinite half-spaces while it 
  takes 
  discrete values inside the tube. 
   
  The scattering matrices of the problem can be constructed 
  by matching the wave functions and their derivatives at the two 
  ends of the cylinder \cite{Zar_Udv,Tekman}. Having obtained the 
  scattering matrices one can proceed by constructing all the 
  scattering states and evaluating the different physical 
  quantities like the LDOS or the conductance of the PC 
  \cite{Zar_Udv}. 
   
  Fig. \ref{fig:ro}.a shows the calculated LDOS fluctuations 
  inside the tube for a PC with $R=15\AA$ and $L=15\AA$ at the 
  point $r=7\AA$ and $z=7\AA$, where 
  the coordinate $z$ is measured from the left wall of the PC. 
  Each time a new conduction channel is opening a peak appears in 
  the function $\varrho(\epsilon,r,z)$. As one can see in Fig. 
  \ref{fig:ro}.b the strong interference effects cause strong 
  fluctuations even for a fixed energy if the spatial coordinate 
  is varied.

  In order to explore the effect of these LDOS fluctuations on the 
  Kondo impurity we construct the next to leading logarithmic 
  scaling equations for an arbitrary LDOS using the multiplicative 
  renormalization group technique \cite{Fowler}. Following 
  Abrikosov \cite{Abrikosov} 
  we describe the impurity spin dynamics by means of a 
  pseudofermion field, $S^i \to b^+_s S^i_{s s^\prime} 
  b_{s^\prime}$, where the operators $b^+_s$ ($b_{s}$) create 
  (annihilate) a pseudofermion corresponding to the spin state 
  $S^{\; z} =s$. The pseudofermion Green's function, 
  ${\cal G}(\omega)$, and the vertex function, 
  $\Gamma_{n n^\prime}(\omega)$ can be introduced in the usual 
  way \cite{Abrikosov}. Due to the special structure of the 
  interaction  the self-energy of the conduction electrons 
   [the vertex function] factorizes: 
  $\Sigma_{n n^\prime}(\omega) = \varphi_n^\ast({\bf r}) 
\Sigma(\omega) 
  \varphi_{n^\prime}({\bf r})$ [ $\Gamma_{n n^\prime}(\omega_i) = 
  \varphi_n^\ast({\bf r}) \Gamma(\omega_i) \varphi_{n^\prime}({\bf 
r})$], 
  where $\Sigma(\omega)$ and $\Gamma(\omega_i)$ 
  depend only on the LDOS at the position of the impurity. 
  Then the multiplicative renormalization group equations can be 
  written in the form \cite{Fowler}: 
  \begin{eqnarray} 
  &&{\cal G}(\omega, j^\prime,D^\prime) 
   = Z (j, D/D^\prime)\;{\cal G}(\omega, j, D) 
  \;, \nonumber \\&& 
  \Gamma(\omega_i, j^\prime, D^\prime) 
   = Z(j, D/D^\prime)^{-1}\; \Gamma (\omega_i, j, D) 
  \;, 
  \label{eq:multipl} 
  \end{eqnarray} 
  where $D^\prime$ is the scaled bandwidth, $Z$ is the 
  pseudofermion wave function renormalization factor, and 
  $j^\prime$ denotes the dimensionless scaled coupling, 
  $j^\prime=J(D^\prime) \varrho_{\rm bulk} (\epsilon_F) $, 
  $\epsilon_F$ being the Fermi energy.

  Similarly to the case of Kondo impurities or two level systems 
  in a smooth LDOS \cite{Fowler,Kondo_rev,Vlad_Zaw} 
  the leading logarithmic terms in the scaling equation arise 
  from the second order vertex corrections while the next 
  to leading logarithmic terms are generated by the second order 
  pseudofermion self energy corrections and the third order vertex 
  corrections. 
  % shown in Fig.~\ref{fig:diagr}. 
  Having calculated these diagrams one can easily generate the 
  scaling equations 
  using Eq. (\ref{eq:multipl}) and after some lengthy algebra one 
  obtains: 
  \begin{eqnarray} 
  {dj  \over dx} &=& j^2  ( R(D) + R(-D)) 
  - 2 j^3 \; \int_0^D {d\xi  \over D} {1 \over (1 + \xi/D)^2} 
  \nonumber \\&& 
  \left(R(\xi)R(-D) + R(-\xi)R(D)\right)  \; , 
  \label{eq:scaling} 
  \end{eqnarray} 
  where $R(\xi)=\varrho(\xi)/\varrho_{\rm bulk}(\epsilon_F)$, 
  $\xi$ being the 
  energy of an electron measured from the Fermi energy, 
  $\xi=\epsilon-\epsilon_F$ and we introduced the 
  scaling variable $x=\ln (D_0/D)$, $D_0\approx\epsilon_F$ being 
  the initial 
  bandwidth cutoff. Naturally, for a system without LDOS 
  fluctuations $R=1$ and Eq.(\ref{eq:scaling}) reduces to the 
  usual next to leading logarithmic scaling equation 
  \cite{Fowler}.

  To see how the LDOS fluctuations  modify the 
  Kondo temperature we drop the next to leading logarithmic term 
  in Eq.(\ref{eq:scaling}) and calculate the leading logarithmic 
  Kondo temperature, which is associated to the divergence of the 
  scaled coupling: $j(D=T_K)=\infty$. Assuming that both the bulk 
  and the fluctuation induced Kondo temperatures, $T_K^*$ and 
  $T_K$ are small enough to satisfy $\varrho(\xi=T_K)\approx 
  \varrho(\xi=T_K^*)\approx \varrho(\epsilon_F)$ one easily 
  obtains the 
  following expression for the ratio of the two Kondo 
  temperatures 
  \begin{equation} 
  {T_K  \over T_K^*} = \exp \left\{ 
  \int_{T^*_K}^D {d\xi \over 2\xi}(\delta R(\xi)+\delta 
  R(-\xi))\right\}\;, 
  \label{eq:ratio} 
  \end{equation} 
  where $\delta R(\xi) = R(\xi)-R_{\rm bulk}(\xi)$. The important 
  feature of Formula (\ref{eq:ratio}) is the appearance of the 
  weight $1/\xi$ in the exponent. The physical meaning of this 
  factor is that the Kondo resonance is mainly formed by the 
  low-energy electron-hole excitations. The contribution of 
  high-energy electron-hole excitations is also significant but it 
  is suppressed compared to that of the low-energy ones. Roughly 
  speaking most of the sites with $\delta\varrho(\epsilon_F) 0$ 
  will have an increased $T_K$ while impurities for 
  which $\delta\varrho(\epsilon_F)<0$ will tend to have a 
  $T_K$ decreased.

  The relative Kondo temperature $T_K/T_K^*$ can be easily 
  estimated by substituting the ratio $\delta R(\xi)$ of the 
  free electron calculations into Eq.(\ref{eq:ratio}). The 
  calculated average ratio $<T_K/T_K^* $ as a function of the 
  radius $R$ of the PC for fixed length $L=5\AA$ is shown in 
  Fig.~2(a). The increased average Kondo temperature, $<T_K \sim 
  1-10 K$, is orders of magnitude larger than bulk Kondo 
  temperature, $T_K^*=0.01K$, and is in reasonable agreement with 
  the experiments (see the diamonds in the Figure) \cite{note}. 
  A  rough fitting of our data with the experimentally found power law 
  dependence $<T_K \sim d^{-\alpha}$  gives an exponent  
$\alpha=2.2\pm 0.5$ 
  which agrees  qualitatively with the experimental exponent, $\alpha 
= 2$.
  (Unfortunately, the available PC diameters were limited by our 
  computer capacity and only a rough estimation of 
  $\alpha$ could have been carried out.) 
  As a comparison we also show the average $<T_K/T_K^* $ 
  calculated by integrating the next to leading logarithmic 
  scaling equations, Eq.(\ref{eq:scaling}). As one can see in 
  Fig.~2(a) there is no significant difference between the next to 
  leading logarithmic and the leading logarithmic results.

  Determining the resistivity contribution of a magnetic impurity 
  at a site $\bf r$ in an ultrasmall PC is a very complicated task 
  because at the length scales involved in the problem 
  the usual Boltzman equation description 
  \cite{phonon_review} breaks down, and presently there is no 
  theory available which could take into 
  account both the geometrical effects ({\it i.e.}, the strong 
  quantum interference) and the strongly correlated behavior of a 
  Kondo impurity in the PC. Therefore we proceed in a 
  semiquantitative way and we estimate the scattering rate 
  $1/\tau(\omega)$  of a conduction electron passing through the 
  orifice. 
  This  is  connected to 
  the imaginary part of the conduction electrons' self energy, 
  $1/\tau(\omega)= -2\; Im \; \Sigma(\omega)$, and  can be 
   measured directly in a PC experiment 
  \cite{Yanson,phonon_review}. Since the 
  conduction electrons' Green's function remains unrenormalized 
  in the dilute impurity limit this scattering rate is simply 
  proportional to $1/\tau(\omega) 
  \sim 2\pi S(S+1) (\varrho(\epsilon_F)J(\omega))^2 \epsilon_F$, 
  where $J(\omega)$  
  denotes the scaled coupling at the energy $D=\omega$. 
   As in the unitary limit the contribution of an impurity in the 
  contact region to the conductance of the PC is approximately 
  $\sim - e^2/h$ in the $\omega=eV\to0$ limit 
  , and   $\varrho(\epsilon_F)J(\omega)\to 1$ for $\omega \to 0$ 
  (see Eq.(\ref{eq:scaling})), the change in the conductance of 
  the PC 
  due to magnetic scattering can be estimated as 
  \begin{equation} 
  \Delta G(eV) = - {e^2 \over h} \Omega\; c  
  \varrho^2(\epsilon_F) \;<J^2(\max\{eV,T\})  \; , 
  \label{eq:cond} 
  \end{equation} 
  where $c$ is the concentration of the impurities, $<... $ 
  denotes the average over the contact region and $\Omega=8R^3/3$ 
  is the effective volume of the PC \cite{Yanson}. 
   
  The calculated amplitudes of the impurity contributions to the 
  zero voltage 
  conductance of the PC as a function of the system size are 
  compared to the experimental data in Fig.~2(b). The average in 
  Eq.~(\ref{eq:cond}) has been carried out over 40 randomly 
  chosen impurity positions. Both the amplitude of $\Delta G(R)$ 
  and 
  its sample to sample fluctuations are in very good agreement 
  with the experiments. We stress at this point that {\it there is 
  no free parameter} in Eq.~(\ref{eq:cond}) exept for the length 
  of the PC which hardly influenced our results. To show that 
  there is a striking size effect in Fig.~2(b) we also plotted the 
  calculated amplitude of the Kondo signal with impurities having 
  the bulk Kondo temperature. Clearly, both the amplitude and the 
  size dependence ($\Delta G\sim d^{2.2}$) of the experimental and 
  our calculated fluctuation dependent  Kondo conductances 
  are quite different from the one we obtained by assuming the 
  bulk Kondo temperature ($\Delta G\sim d^3$). 
   
  The size dependence of the Kondo resonance can be understood as 
  follows. Decreasing the contact size strong LDOS fluctuations 
  evolve in the contact region. The smaller the contact size the 
  larger the fluctuations become, and thus for small contact sizes 
  the measured few impurities (i.e., the impurities in the contact 
  region) have a very broad distribution of Kondo temperatures. 
  Since in the temperature range $\sim 1 K$ where the 
  measurements were performed \cite{Yanson} magnetic impurities 
  with a small Kondo temperature, $T_K\approx T_K^* \sim 0.01 K$, 
  give 
  no significant contribution to Eq.(\ref{eq:cond}) the 
  differential 
  conductance  is always dominated by the impurities with 
  the largest Kondo temperatures.  These 
  arguments suggest that the sum in Eq.(\ref{eq:cond}) is 
  dominated by the magnetic impurities having  large Kondo 
  temperatures. Therefore our 
  theory gives a natural explanation why the applied magnetic 
  field could not destroy the zero bias Kondo anomaly for small 
  samples \cite{Yanson}.

  One can also show analyzing formula (\ref{eq:ratio}) that for a 
  fixed PC shape the relative increase of the Kondo temperature, 
  $T_K/T_K^*$, depends essentially on the bulk Kondo temperature 
  $T_K^*$, and for alloys with very small $T_K^*$ the ratio 
  $T_K/T_K^*$ can be much larger than for alloys with large 
  $T_K^*$. 
  This explains why no significant broadening  of the Kondo peak 
  has 
  been observed in $CuFe$ PC's $(T_K^* \approx 20 K)$ 
  \cite{Naidyuk}, while an enormous broadening was found in $AuMn$ 
  and $CuMn$ ($T_K^* < 0.01 K$) \cite{Jansen,Yanson}. 
   
  According to the picture proposed in this Letter one expects 
  that any physical quantity which is sensitive on the LDOS will 
  have a distribution in the contact region due to the presence of 
  strong fluctuations. As an example we mention the PC spectrum of 
  fast 
  two level systems (TLS) \cite{Ralph_Ludwig,Keijsers}.  Since the 
  dimensionless 
  TLS -- conduction electron couplings are sensitive on the LDOS 
  \cite{Vlad_Zaw,Zar_Zaw} and $T_K^*$ is in the range of $\sim 1K$ 
  \cite{Ralph_Ludwig} we expect a slight broadening of the zero 
  bias TLS peak, which has indeed been observed in metallic break 
  junctions \cite{Keijsers}. 
   
  In our simple model the LDOS fluctuations were basically 
  connected to the openings of new conductance channels through 
  the PC 
  \cite{Zar_Udv}. 
  Such LDOS fluctuations, however, may also be generated by 
  random scattering at 
  the boundary of the PC. Effectively, it has been found recently 
  that random scattering at the surface of a PC is able to 
  produce huge LDOS fluctuations in the contact region 
  \cite{Kozub}. Therefore 
  we think that our conclusions are qualitatively independent of 
  the special 
  model considered. 
   
  The case of ultrasmall PC's with magnetic impurities 
  \cite{Yanson} should be contrasted to the 
  experiments performed on thin metallic films and wires 
  \cite{Kondo_films}. 
  In the latter case only alloys with high Kondo 
  temperatures ($CuFe$, $AuFe$) were investigated, where the LDOS 
  fluctuations do 
  not modify the Kondo temperature very much. Instead, it seems to 
  be that spin orbit interaction induced spin anisotropy develops 
   which leads to the blocking of the spin flip 
  processes \cite{Ujs_Zawa} and a suppression of the Kondo signal. 
  On the contrary for thin $AuMn$ films, where $T_K^*$ is very 
  small, we expect that the Kondo temperature can be increased due 
  to the surface induced LDOS fluctuations 
  by several orders of magnitude similarly to the case of PC's. 
   
  In summary, we have proposed that the recently measured 
  anomalous Kondo temperatures in small $CuMn$ PC's are due to the 
  strong LDOS fluctuations in the contact region generated by the 
  scattering of the conduction electrons at the surface of the PC. 
  Carrying out a model calculation we found that  these 
  fluctuations are really able to increase the Kondo temperature 
  by several orders of magnitude in agreement with the 
  measurements. 
  Our results also suggest that the Boltzman equation approach 
  becomes inadequate for ultrasmall PC spectrometry, where quantum 
  interference and fluctuation effects may become extremely 
  important.

  The authors are deeply indebted to A. Zawadowski, I. K. Yanson, 
  Jan van Ruitenbeek, N van der Post, and I. V. Kozub for helpful 
  discussions. This research has been supported by the Hungarian 
  Grants OTKA~7283/93 and OTKA~F016604 . 
  %\widetext 

  \begin{figure}[tb] 
  %\epsfxsize=7truecm 
  %\hskip0.5truecm\epsfbox{ujabra.eps}\hfill 
  %\vskip0.1truecm 
  \caption{\label{fig:ro} 
  (a): Fluctuations of the LDOS $\varrho(\epsilon,z,r)$ inside the 
  tube 
  for a PC with $R=15\AA$ and $L=15\AA$ at the 
  point $r=7\AA$ and $z=7\AA$. The fluctuating LDOS does not 
  integrate exactly to its bulk value (dashed line) because the 
  hard wall of the PC pushes the conduction electrons in the 
  inside region of the tube. (b): $\varrho(\epsilon,z,r)$ for 
  the same PC at $z=7\AA$ and energy $\epsilon=7 eV$ as a function 
  of the radius $r$.} 
  \end{figure}  
   
  \begin{figure}[htb] 
  %\epsfxsize=7truecm 
  %\hskip0.5truecm\epsfbox{g-tk.eps} 
  %\vskip0.1truecm 
  \caption{\label{fig:sizedep} 
  (a) The average relative Kondo temperatures $<T_K/T_K^* $ as a 
  function of the radius $R$ of a $CuMn$ PC with $L=5 \AA$. The 
  cutoff and the bare couplings have been chosen to be 
  $D_0=6.8eV$ and $j_0=0.032$. Crosses and diamonds denote the 
  results obtained in the leading and in the next to leading 
  logarithmic orders, while data points indicated by boxes have 
  been 
  obtained from the results of Ref.~\protect{\cite{Yanson}}. 
  The large fluctuations are due to the sensitivity of the 
  interference pattern to the geometry of the PC. 
  (b) Size dependence of the amplitude of the dimensionless Kondo 
  conductance $\Delta g=\Delta G\; h/e^2$ 
  for the same contact. Diamonds denote the experimental data 
  taken from Ref.~\protect{\cite{Yanson}} while our results are 
  indicated by crosses. The impurity concentration and the 
  temperature were $c=0.1\%$ and $T=0.05K$, respectively. The 
  dashed line indicates the results without LDOS fluctuations ($g 
  \sim R^{3}$) 
  while the continuous line corresponds to the best fit to the 
  data of Ref.~\protect{\cite{Yanson}}: $g\sim 
  R^{2.17}$.} 
  \end{figure}  
   

\begin{references} 
   
  \bibitem{Lysykh} A. A. Lysykh, I. K. Yanson, O. I. 
  Shklyarevskii, 
  and Yu. G. Naidyuk, Solid State Commun. {\bf 35}, 987 (1980). 
   
  \bibitem{Omel} A. N. Omel'yanchuk and I. G. Tuluzov, Fiz. Nizk. 
  Temp.,  {\bf 6}, 1286 (1980) [Sov. J. Low. Temp. Phys., {\bf 6} 
  626 (1980)]. 
   
  \bibitem{Jansen} A. G. M. Jansen, A. P. van Gelder, P. Wyder and 
  S. Str{\" a}ssler, J. Phys. F: Metal Phys. {\bf 11}, L15 (1981). 
   
  \bibitem{Naidyuk} Yu. G. Naidyuk, O. I. Shklyarevskii, and I. K. 
  Yanson, Fiz. Nizk. Temp. {\bf 8}, 725 (1982) [Sov. J. Low. Temp. 
  Phys. {\bf 8}, 362 (1982)]. 
   
  \bibitem{Yanson} I.K. Yanson {\it et al.}, Phys. Rev. Lett. {\bf 
  74}, 302 (1995); I.K. Yanson {\it et al.}, Low Temp. Phys. {\bf 
  20}, 836 (1994). 
   
  \bibitem{Kondo_films} G. Chen and N. Giordano, Phys. Rev. Lett. 
  {\bf 66}, 209 (1991); J. F. DiTusa {\it et al.}, Phys. Rev. 
  Lett. {\bf 68}, 678 (1992). 
   
  \bibitem{Dobros} V. Dobrosavljevi\'c, T.R. Kirkpatrick, and G. 
  Kotliar, Phys. Rev. Lett. {\bf 69}, 1113 (1992). 
   
  \bibitem{supers} K. B. Efetov and V. N. Prigodin, Phys. Rev. 
  Lett. {\bf 70}, 1315 (1993); 
  A. D. Mirlin and Y. V. Fyodorov, Phys. Rev. Lett. {\bf 72}, 526 
  (1994). 
   
  \bibitem{Zar_Udv} G. Zar\'and and L. Udvardi,  Physica B {\bf 
  218}, 68 (1996). 
   
  \bibitem{Tekman} E. Tekman and S. Ciraci, Phys. Rev. B {\bf 39}, 
  8772 (1989). 
   
   
  \bibitem{Fowler} M. Fowler and A. Zawadowski, Solid State 
  Commun. {\bf 9}, 471 (1971). 
   
  \bibitem{Abrikosov} A. A. Abrikosov, Physics {\bf 2}, 5 (1965). 
  %pseudoferm 
   
  \bibitem{Kondo_rev} G. Gr{\" u}ner, Adv. in Phys. {\bf 23}, 941 
  (1974); G. Gr\"uner and A. Zawadowski, Rep. Prog. Phys. {\bf 
  37}, 1500 (1974). 
  %green review 
   
  \bibitem{phonon_review} I.K. Yanson and O. I. Shklyarevskii, 
  Fiz. Nizk. Temp. {\bf 12}, 899 (1986) [Sov. J. Low Temp. Phys. 
  {\bf 12}, 509 (1986); A. G. Jansen, A. P. van Gelder and P. 
  Wyder, J. Phys. C {\bf 13}, 6073 (1980). 
   
  \bibitem{Ralph_Ludwig} See D.C.\ Ralph, A.W.W.\ Ludwig, Jan von 
  Delft, and 
  R.A. Buhrman, Phys.\ Rev.\ Lett.\ {\bf72}, 1064 (1994);D.C.\ 
  Ralph and R.A.\ Buhrman, Phys.\ Rev.\ Lett.\ {\bf69}, 2118 
  (1992). 
   
  %two-channel scaling 
   
  \bibitem{Keijsers} R. J. P. Keijsers, O. I.  Shklyarevskii, and 
  H. 
  van Kempen, Phys. Rev. B {\bf 51}, 5628 (1995). 
   
  \bibitem{Vlad_Zaw} K.\ Vlad\'ar and A.\ Zawadowski, Phys.\ Rev.\ 
  B {\bf28}, 
  1564, 1582, 1596 (1983). 
   
  \bibitem{Zar_Zaw} G.\ Zar\'and and A.\ Zawadowski, 
  Phys.\ Rev.\ Lett.\ {\bf72}, 542 (1994). 
   
  \bibitem{note} Since the weak coupling analysis is inappropriate 
  for very small PC's (see Ref.~\protect{\cite{Yanson}}) for very 
  small contact diameters $d$ corresponding to our calculations we 
  extracted the Kondo temperatures from the data of 
  Ref.~\cite{Yanson} by measuring the width of the Kondo resonance 
  curves. 
   
  \bibitem{Kozub} V. I. Kozub (private communication). 
   
  \bibitem{Ujs_Zawa} O. \'Ujs\'aghy and A. Zawadowski 
  (unpublished). 
   
   
  \end{references}
  \end{document}